\documentclass[twocolumn,showpacs,preprintnumbers,amsmath,amssymb]{revtex4}

\usepackage{graphicx}
\usepackage{dcolumn}
\usepackage{bm}

\begin{document}
\newcommand{\Arg}[1]{\mbox{Arg}\left[#1\right]}
\newcommand{\bb}{\mathbf}
\newcommand{\braopket}[3]{\left \langle #1\right| \hat #2 \left|#3 \right \rangle}
\newcommand{\braket}[2]{\langle #1|#2\rangle}
\newcommand{\be}{\[}
\newcommand{\br}{\vspace{4mm}}
\newcommand{\bra}[1]{\langle #1|}
\newcommand{\braketbraket}[4]{\langle #1|#2\rangle\langle #3|#4\rangle}
\newcommand{\braop}[2]{\langle #1| \hat #2}
\newcommand{\dd}[1]{ \! \! \!  \mbox{d}#1\ }
\newcommand{\DD}[2]{\frac{\! \! \! \mbox d}{\mbox d #1}#2}
\renewcommand{\det}[1]{\mbox{det}\left(#1\right)}
\newcommand{\ee}{\]} 
\newcommand{\eg}{\textbf{\\  Example: \ \ \ }}
\newcommand{\Imag}[1]{\mbox{Im}\left(#1\right)}
\newcommand{\ket}[1]{|#1\rangle}
\newcommand{\ketbra}[2]{|#1\rangle \langle #2|}
\newcommand{\kp}{\arccos(\frac{\omega - \epsilon}{2t})}
\newcommand{\ldos}{\mbox{L.D.O.S.}}
\renewcommand{\log}[1]{\mbox{log}\left(#1\right)}
\newcommand{\Log}{\mbox{log}}
\newcommand{\Modsq}[1]{\left| #1\right|^2}
\newcommand{\nb}{\textbf{Note: \ \ \ }}
\newcommand{\op}[1]{\hat {#1}}
\newcommand{\opket}[2]{\hat #1 | #2 \rangle}
\newcommand{\occ}{\mbox{Occ. Num.}}
\newcommand{\Real}[1]{\mbox{Re}\left(#1\right)}
\newcommand{\so}{\Rightarrow}
\newcommand{\sol}{\textbf{Solution: \ \ \ }}
\newcommand{\thetafn}[1]{\  \! \theta \left(#1\right)}
\newcommand{\tin}{\int_{-\infty}^{+\infty}\! \! \!\!\!\!\!}
\newcommand{\Tr}[1]{\mbox{Tr}\left(#1\right)}
\newcommand{\kb}{k_B}
\newcommand{\rad}{\mbox{ rad}}
\preprint{APS/123-QED}

\title{Electronic structure of graphene beyond the linear dispersion regime}

\author{S. R. Power and M. S. Ferreira}
\email{ferreirm@tcd.ie}
\affiliation{
School of Physics, Trinity College Dublin, Dublin 2, Ireland}

\date{\today}

\begin{abstract}
Among the many interesting features displayed by graphene, one of the most attractive is the simplicity with which its electronic structure can be described. The study of its physical properties is significantly simplified by the linear dispersion relation of electrons in a narrow range around the Fermi level. Unfortunately, the mathematical simplicity of graphene electrons is only limited to this narrow energy region and is not very practical when dealing with problems that involve energies outside the linear dispersion part of the spectrum. In this communication we remedy this limitation by deriving a set of closed-form analytical expressions for the real-space single-electron Green function of graphene which is valid across a large fraction of the energy spectrum. By extending to a wider energy range the simplicity with which graphene electrons are described, it is now possible to derive more mathematically transparent and insightful expressions for a number of physical properties that involve higher energy scales. The power of this new formalism is illustrated in the case of the magnetic (RKKY) interaction in graphene.

\end{abstract}
                 
\maketitle
\bibliographystyle{abbrv} 

\section{Introduction}
Graphene-related materials have been in the scientific limelight for the past few years due to the numerous applications envisaged for them\cite{riseofgraphene}. Besides the huge potential for applicability, one key feature that makes graphene particularly popular is the simplicity with which many of its physical properties can be described, primarily due to the simple dispersion relation for its electrons. The linearity of this dispersion relation around the Fermi level enables the  description of graphene electrons in terms of massless Dirac fermions\cite{RMP-AHCN}. This introduces a great level of mathematical transparency in the portrayal of their properties, which nevertheless is limited only to a narrow range of energies around the Fermi level. Energy values outside this range are often needed, for example when gated graphene systems are considered\cite{riseofgraphene} or when calculation of a relevant physical quantity requires an integral over energy, but lack the mathematical transparency of those within the linear dispersion regime. 

In this communication we show how this limitation can be circumvented by deriving a fully analytical closed-form expression for the single-electron Green function of graphene in real space, a quantity that is instrumental in describing the behaviour of graphene electrons. Because Green functions are used in the study of several physical properties, improvements in their mathematical description will enable far more transparent and insightful expressions for the corresponding physical quantities. This is particularly true for distant-dependent interactions across a graphene sheet, for example the effect of an impurity on the physical properties of the system at a certain distance away from where it is introduced. Another example is the interaction of two impurities embedded into the sheet. In both cases, for distances of more than a few lattice spacings, the interactions involved are mediated by the conduction electrons of the graphene host. The Green functions calculated in this paper describe the equilibrium properties of these electrons at low temperatures, allowing us to investigate the underlying interactions within a mathematically transparent framework. Such a methodology allows for the prediction of certain features of the interaction without recourse to numerical calculations.

The remainder of the paper is organised as follows. The general method for calculating the Green function required is introduced in Section \ref{methods_sec}. The important directions of the graphene geometry, namely the \emph{armchair} and \emph{zigzag} directions are illustrated and an explicit calculation for the real-space off-diagonal Green function element in each of these directions is performed in sections \ref{arm_dir} and \ref{zig_dir}, respectively. The accuracy of our approach is demonstrated by comparison with a fully numerical calculation. An extension of the method for arbitrary directions and inter-sublattice cases is discussed in section \ref{oth_dir}. The potential applications of our approach are discussed in Section \ref{gen_app}, before an explicit illustration is given for the case of the magnetic interaction in graphene in section \ref{rkky_app}. Here a fully analytical method is used to derive the principal features of the interaction within the RKKY approximation. 

\section{Method and Calculations}
\label{methods_sec}
The general formula for the single-electron Green function, within the nearest-neighbour tight-binding framework, in its eigenstate basis is 
\begin{equation}
{\hat {\cal G}}(E )=\sum_{\vec k} \left\{ \frac{\vert {\vec k},+ \rangle
\langle {\vec k},+ \vert}{E -{\cal E}_{+}({\vec k})}+ \frac{ \vert {\vec
k},- \rangle \langle {\vec k},- \vert}{E -{\cal E}_{-}({\vec
k})}\right\}\,,
\label{General}
\end{equation}
where $E$ is the energy, $\vert {\vec k},\pm \rangle$ is the eigenvector labeled by the wave vector ${\vec k}$ and ${\cal E}_\pm $ is the corresponding eigenvalue defined as
\begin{equation}
{\cal E}_{\pm}({\vec k})=\pm t \sqrt{1+4\cos(\frac{\sqrt{3}k_y\,a} {2})\cos
(k_x\frac{a}{2})+4\cos {}^{2}(k_x\frac{a}{2})} \,\,.
\label{graphene}
\end{equation} 

\begin{figure}
\includegraphics[width=0.3\textwidth]{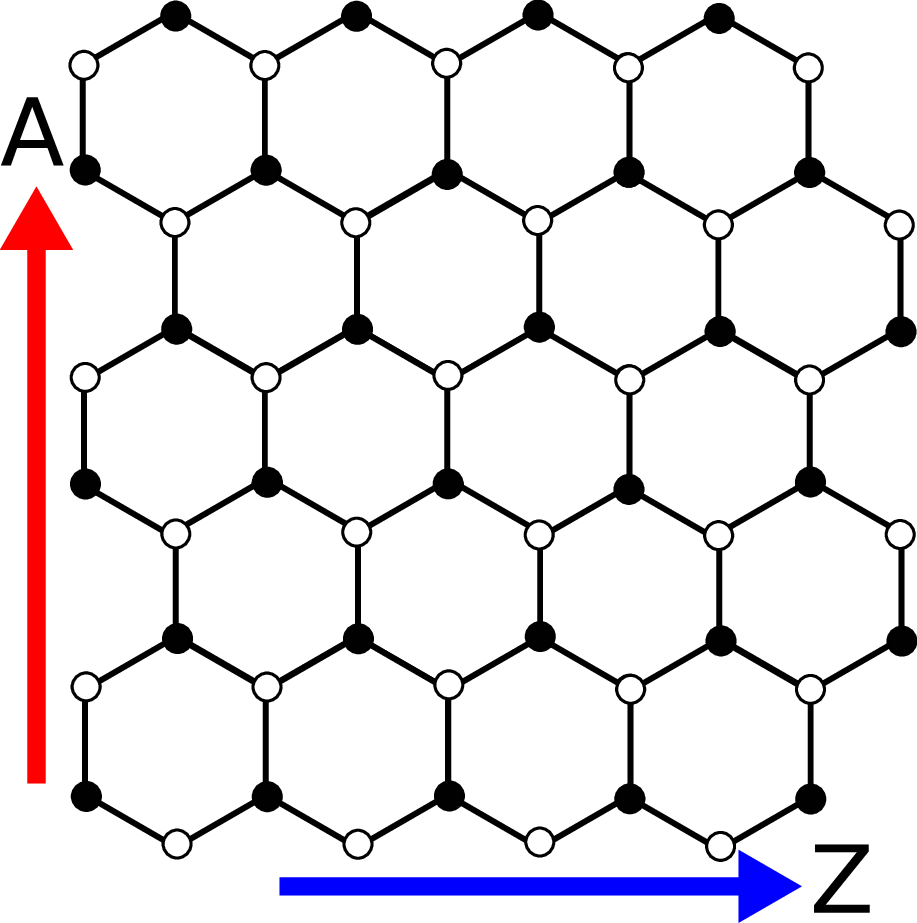}
\caption{Schematic of the graphene lattice with the zigzag (armchair) direction denoted by the blue (red) arrow marked `Z' (`A'). The filled and hollow circles represent carbon atoms on each of the two sublattices which compose the graphene lattice.}

\label{schematic}
\end{figure}
The quantities $a$ and $t$ correspond to the lattice parameter of graphene and its nearest-neighbour electronic hopping\cite{hopping_note}, respectively, which are hereafter used as our units of distance and energy. Our choice is such that the $x$ ($y$) direction is aligned to the zigzag (armchair) geometry of the graphene lattice as is shown schematically in Fig. \ref{schematic}. A simple unitary transformation defines the real-space basis $\vert j,\zeta \rangle$, where the index $j$ labels the two-atom unit cell and the index $\zeta$ refers to the intra-cell atoms corresponding to the two distinct sublattices of graphene, represented by filled and hollow circles in Figure \ref{schematic}. When projected onto two different states $\vert j,\zeta \rangle$ and $\vert j^\prime,\zeta \rangle$ located in real space by the respective vectors ${\vec R}_j$ and  ${\vec R}_{j^\prime}$, the Green function is written as 
\begin{equation}
\langle j, \zeta \vert {\hat {\cal G}}(E) \vert j^\prime, \zeta \rangle=\frac{a^2 \sqrt{3}}{8 \pi^2 }\int dk_x \, \int dk_{y} \, \frac{E \,\,e^{i {\vec k}.({\vec R}_j - {\vec R}_{j^\prime})} }{E^{2}-{\cal E}_{+}^{2}({\vec k})}\,,
\label{above}
\end{equation}
where the integrals over $k_x$ and $k_y$ are performed over the first Brillouin Zone of graphene.
Although we have selected two states that belong to the same sublattice ($\zeta = \zeta^\prime$), this constraint can be easily relaxed and generalized to describe the propagator between sites on different sublattices. Before proceeding, we will outline the basic steps taken to obtain the Green function. We tackle the first integral by analytical continuation to the complex plane, where it is subsequently solved using the residue theorem.\cite{kirwan1, boasbook} The remaining integral can then be solved in the case of moderately large separation vectors by using the Stationary Phase Approximation (SPA).\cite{spa}

\subsection{Armchair Direction}
\label{arm_dir}
\begin{figure}
\includegraphics[width=0.4\textwidth]{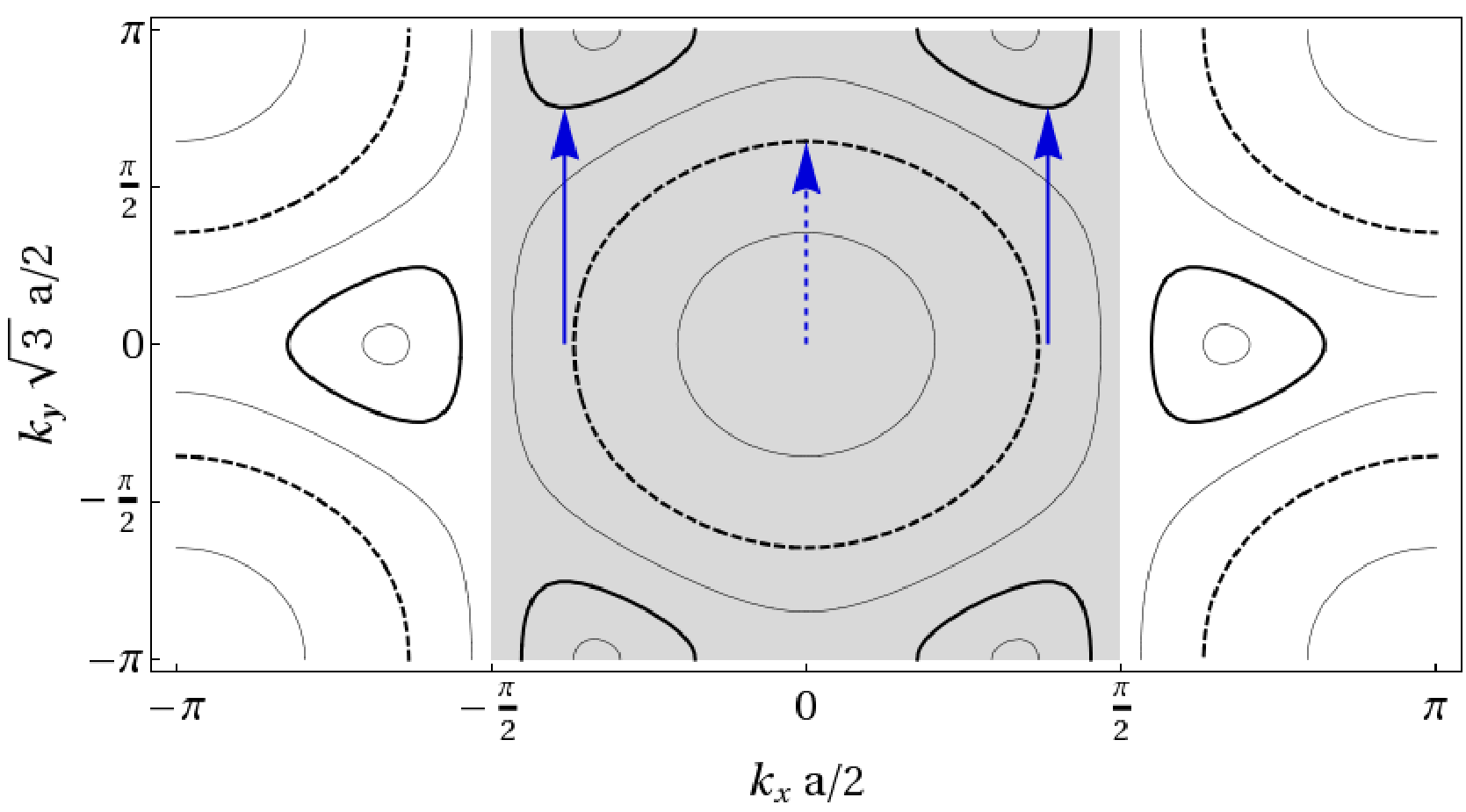}
\caption{Constant energy plots of the function ${\cal E}_+({\vec k})$ in reciprocal space for a few different energies. Horizontal and vertical axes are rescaled as $k_x a/2$ and $k_y a \sqrt{3}/2$, respectively, so that they are plotted as dimensionless quantities. The rectangular shaded area delimits the first Brillouin zone over which the integrals for the armchair direction case are taken. Constant energy plots for two specific energies, $E=0.7 |t|$ (solid) and $E=1.8 |t|$ (dashed), are drawn with thicker lines. The corresponding stationary wave vectors ${\tilde q}$ for these energies are highlighted with arrows. Note that the sign of ${\tilde q}$, shown as positive here for simplicity, must be chosen according to the conditions outlined in the text.}
\label{constant_energy}
\end{figure}

We first consider the case of separation vectors ${\vec R}_j - {\vec R}_{j^\prime}$ along the (armchair) $y$-direction. By showing how to obtain the Green function for this particular case we hope to illustrate the general method for calculating Green functions for any direction. Note that the position vectors appear only as a difference and can be further simplified by defining $\Delta \equiv \vert {\vec R}_j - {\vec R}_{j^\prime}\vert$. In this case, the integral is performed over the Brillouin Zone shown in Fig. \ref{constant_energy} and the first integral, over $k_y$, can be evaluated by extending $k_{y}$ to the realm of complex numbers and changing the integration contour from a straight line on the real axis to the boundaries of a semi-infinite rectangle in the upper half of the complex plane, with its base lying on the real axis between $-{2\pi \over a \sqrt{3}}$ and ${2\pi \over a \sqrt{3}}$. Because the integrand vanishes in the limit ${\rm Im}[k_y] \rightarrow \infty$ and because the parts of the contour that are parallel to the imaginary axis cancel each other out, the $k_y$-integral can be evaluated by simply identifying the poles of the integrand lying inside the integration contour and finding their respective residues \cite{kirwan1}, that is, 
\begin{equation}
{\cal G}_{\Delta}(E) = {i \, a \over 4 \pi t^2} \int_{-\pi \over a}^{\pi\over a} \, dk_x \, {E \, e^{i q \Delta} \over \cos({k_x \, a \over 2}) \, \sin({q a \sqrt{3} \over 2})}\,\,.
\label{remaining}
\end{equation} 
Note that the scalar product $\langle j, \zeta \vert {\hat {\cal G}}(E) \vert j^\prime, \zeta^\prime \rangle$ is now more concisely expressed as ${\cal G}_{\Delta}(E)$ and that 
\begin{equation}
\cos({q a \sqrt{3} \over 2}) = {{E^2 \over t^2} - 1 - 4 \cos^2({k_x a \over 2}) \over 4 \cos({k_x a \over 2})}
\label{poles}
\end{equation}
defines the wave vector $q$ that comes out of the first integral. Although Eq.(\ref{poles}) provides a unique definition for $\cos({q a \sqrt{3} \over 2})$, it does not specify the sign of $q$ uniquely. Its sign is selected by imposing that $q$ must necessarily lie within the integration contour of the $k_y$-integral. 

The $k_x$-dependence contained in the wave vector $q$, as seen in Eq.(\ref{poles}), means that the integrand in Eq. (\ref{remaining}) is an oscillatory function of $k_x$ that oscillates very rapidly for large values of the separation $\Delta$. In this case, the only non-vanishing contribution to the Green function comes from regions for which the phase of the exponential function is stationary. To locate these stationary points we must impose that $dq/dk_x=0$, which leads to solutions of the form 
\begin{equation}
{\tilde k}_x = \left\{ 
\begin{array}{l l}
  \pm { 2 \over a} \, \cos^{-1}\left({\sqrt{t^2 - E^2} \over 2 t  }\right) & \quad \mbox{if $\vert E \vert < \vert t \vert$}\\
  0 & \quad \mbox{if $\vert E \vert > \vert t \vert$}\\ \end{array} \right.
\label{tilde}  
\end{equation}
Note that due to a topological change in the constant energy surfaces of the function ${\cal E}_+({\vec k})$ at $E= \pm t$ we separate the energy band into two separate regions, namely, the inner region defined by $\vert E \vert < \vert t \vert$ and the outer region defined by $\vert E \vert > \vert t \vert$. Both stationary solutions written above are valid throughout the entire energy spectrum. However, outside those specified regions, albeit solutions, they give rise to complex values for the wave vectors $q$. With complex wave vectors, the integrand of Eq.(\ref{remaining}) tends to vanish for any sizable separation $\Delta$, meaning that the stationary values outside the ranges listed in Eq.(\ref{tilde}) should have very little influence in the overall results for the Green functions.\cite{obs} The same can be understood from purely geometrical arguments applied to the constant energy surfaces of ${\cal E}_+({\vec k})$ in reciprocal space, depicted in Fig. \ref{constant_energy}.  When searching for stationary solutions for $q$, which in this case lie parallel to the $y$-axis, the two solutions resulting from Eq.(\ref{tilde}) are the only possible (real) ones within the rectangular Brillouin zone of the hexagonal lattice. The tilde symbol ($\sim$) will hereafter be used to refer to the values of $k_x$ and $q$ satisfying the stationary condition. Therefore, the wave vector $q$ when expanded in a Taylor series around the stationary value ${\tilde k}_x$ has no linear component and, up to second order, is approximated by 
\begin{equation}
q \approx C_1 + C_2 \left(k_x - {\tilde k}_x\right)^2\,\,,
\label{q}
\end{equation}
where

\begin{equation}
C_1 = \left\{ 
\begin{array}{l l}
  \pm {2 \over a \sqrt{3}} \, \cos^{-1}\left({-\sqrt{t^2-E^2}\over t}\right) & \quad \mbox{if $\vert E \vert < \vert t \vert$}\\
  \pm {2 \over a \sqrt{3}} \, \cos^{-1}\left({E^2 - 5t^2 \over 4t^2}\right) & \quad \mbox{if $\vert E \vert > \vert t \vert$}\\ \end{array} \right.
\label{c1}  
\end{equation}

and

\begin{equation}
C_2 = \left\{ 
\begin{array}{l l}
  \pm {a \over 4 \sqrt{3}} \, \left({E^2+3t^2 \over E \sqrt{t^2-E^2}}\right) & \quad \mbox{if $\vert E \vert < \vert t \vert$}\\
  \pm {a \over 4 \sqrt{3}} \, \left({E^2+3t^2 \over \sqrt{(t^2-E^2)(E^2 - 9t^2}}\right) & \quad \mbox{if $\vert E \vert > \vert t \vert$} \,. \\ \end{array} \right.
\label{c2}  
\end{equation}

Note that the sign of $C_1$ must be chosen to ensure that $q$ lies within the $k_y$ integration contour as before. The sign of $C_2$ is determined by its correspondence to the curvature of $q$ at ${\tilde k}_x$.
If we now insert Eq.(\ref{q}) into Eq.(\ref{remaining}) and make use of the fact that $k_x$ and $q$ will not vary very much around their respective stationary values ${\tilde k}_x$ and ${\tilde q}$, we are left with a much simplified expression for the Green function, which now reads
\begin{equation}
{\cal G}_{\Delta}(E) = {i a E e^{i C_1 \Delta} \over 4 \pi t^2 \cos\left({{\tilde k}_x a \over 2}\right) \sin\left({{\tilde q} a \sqrt{3} \over 2}\right)} \int d k_x \, e^{i C_2 (k_x - {\tilde k}_x)^2 \Delta} \,\,.
\end{equation}
The remaining integral is a well known Gaussian integral whose solution gives
\begin{equation}
{\cal G}_{\Delta}(E) = {i a E e^{i C_1 \Delta} \over 4 \pi t^2 \cos\left({{\tilde k}_x a \over 2}\right) \sin\left({{\tilde q} a \sqrt{3} \over 2}\right)} \sqrt{{i \, \pi \over C_2 \, \Delta}}\,\,.
\label{sol_x}
\end{equation}

This can be rewritten in a more transparent fashion using the definitions in Eqs. (\ref{tilde}), (\ref{c1}) and (\ref{c2}) to provide a completely analytical expression for the off-diagonal Green function matrix element between two graphene sites separated by a distance $\Delta$ along the armchair direction. For positive values of energy ($E > 0$), we find

\begin{figure}
\includegraphics[width=0.4\textwidth]{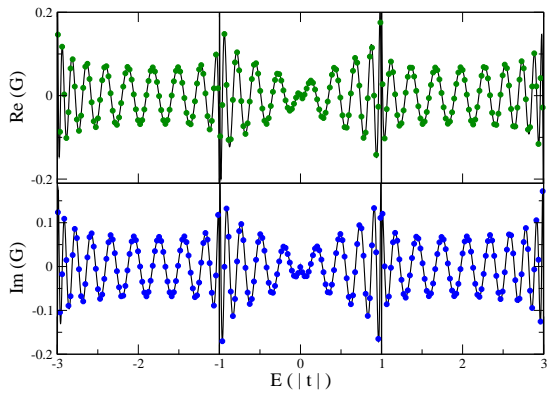}
\caption{${\cal G}_{\Delta}(E)$ as a function of energy (in units of $|t|$) for the case of $\Delta = 10\sqrt{3}a$. Top (Bottom) panel shows the real (imaginary) part of the Green function. Lines correspond to the results evaluated by Eq.(\ref{sol_x}) whereas points are the result of brute-force numerical calculations of Eq.(\ref{above}). }
\label{figarm}
\end{figure}

\begin{widetext}
\begin{equation}
 {\cal G}_{\Delta}(E)   = \sqrt{\frac{2}{i \pi}} {{1} \over {\sqrt{(E^2 + 3t^2) \sqrt{t^2-E^2}}}} \sqrt{\frac{1}{\Delta^{\prime}}}  \left\{ 
\begin{array}{l l}
   -i \sqrt{E} (\frac{i E + \sqrt{t^2-E^2}}{t})^{\Delta^{\prime}}& \quad \mbox{if $ E  < \vert t \vert$}\\
   \frac{E}{\sqrt{\sqrt{E^2 -9t^2}}} (\frac{E^2 -5t^2 - i \sqrt{(t^2-E^2)(E^2-9t^2)}}{4t^2})^{\Delta^{\prime}} & \quad \mbox{if $ E > \vert t \vert$ \,\,,}\\ \end{array} \right.
\label{gf_et}  
\end{equation}
\end{widetext}
where $\Delta^{\prime} = \frac{2 \Delta}{a \sqrt{3}}$. 
We again consider the distinct cases $E < \vert t \vert$ and $E > \vert t \vert$ and note that the only occasion when both stationary points contribute is at energies very near $E = \pm t$ \cite{obs}.
Fig. \ref{figarm} compares both the real and imaginary parts of the Green function for the case of $\Delta = 10 \sqrt{3} a$ obtained by the analytical expression above with those obtained through a numerical evaluation of Eq.(\ref{above}). For $E<0$ we note that the real (imaginary) part of the Green function is an odd (even) function of energy, as can be seen in Fig. \ref{figarm}, and make use of the relations $\mathrm{Re}({\cal G}_{\Delta}(-E)) = - \mathrm{Re}({\cal G}_{\Delta}(E))$ and $\mathrm{Im}({\cal G}_{\Delta}(-E)) =  \mathrm{Im}({\cal G}_{\Delta}(E))$.

\begin{figure}
\includegraphics[width=0.4\textwidth]{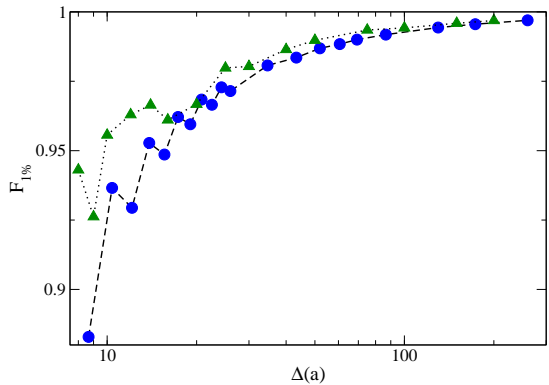}
\caption{Fraction ${\cal F}_{1\%}$ of the bandwidth for which the error between the analytical and brute force numerical results is below 1\%, plotted as a function of the separation $\Delta$ (in units of $a$). The circles (triangles) correspond to separations in the armchair (zigzag) direction.}
\label{figerror}
\end{figure}

At first glance one might think that the replacement of the well established linear dispersion approximation with another that is valid only for asymptotically large values of separation is unlikely to improve the range of validity of the overall result. However, as seen in Fig. \ref{figarm} there is hardly any noticeable difference between the analytical and numerical results across the entire energy band. Because the analytical expression relies on the stationary phase argument, this remarkable agreement is likely to regress as the separation ($\Delta$) is reduced. 
To test how good an approximation Eq.(\ref{sol_x}) is, in Fig. \ref{figerror} the fraction ${\cal F}_{1\%}$ of the bandwidth for which the relative error between the numerical and analytical evaluations is less than $1\%$ is plotted as a function of the separation. The plot with circular points corresponds to the armchair direction. Even for small separations ($\Delta \approx 10a$), the energy range for which the Green functions are very accurately described by our analytical expression exceeds 90\% of the bandwidth. In other words, there is only a very narrow energy range in which the disagreement exceeds $1\%$. Most remarkably, as the separation is increased this small range decreases very rapidly, indicating that Eq.(\ref{sol_x}) is capable of accurately describing the Green function across almost the entire energy spectrum for separation values larger than a few lattice parameters.  This is a major advantage when compared to the narrow fraction of the bandwidth that meets the linear dispersion criterion. 

\subsection{Zigzag direction}
\label{zig_dir}
\begin{figure}
 \includegraphics[width=0.45\textwidth]{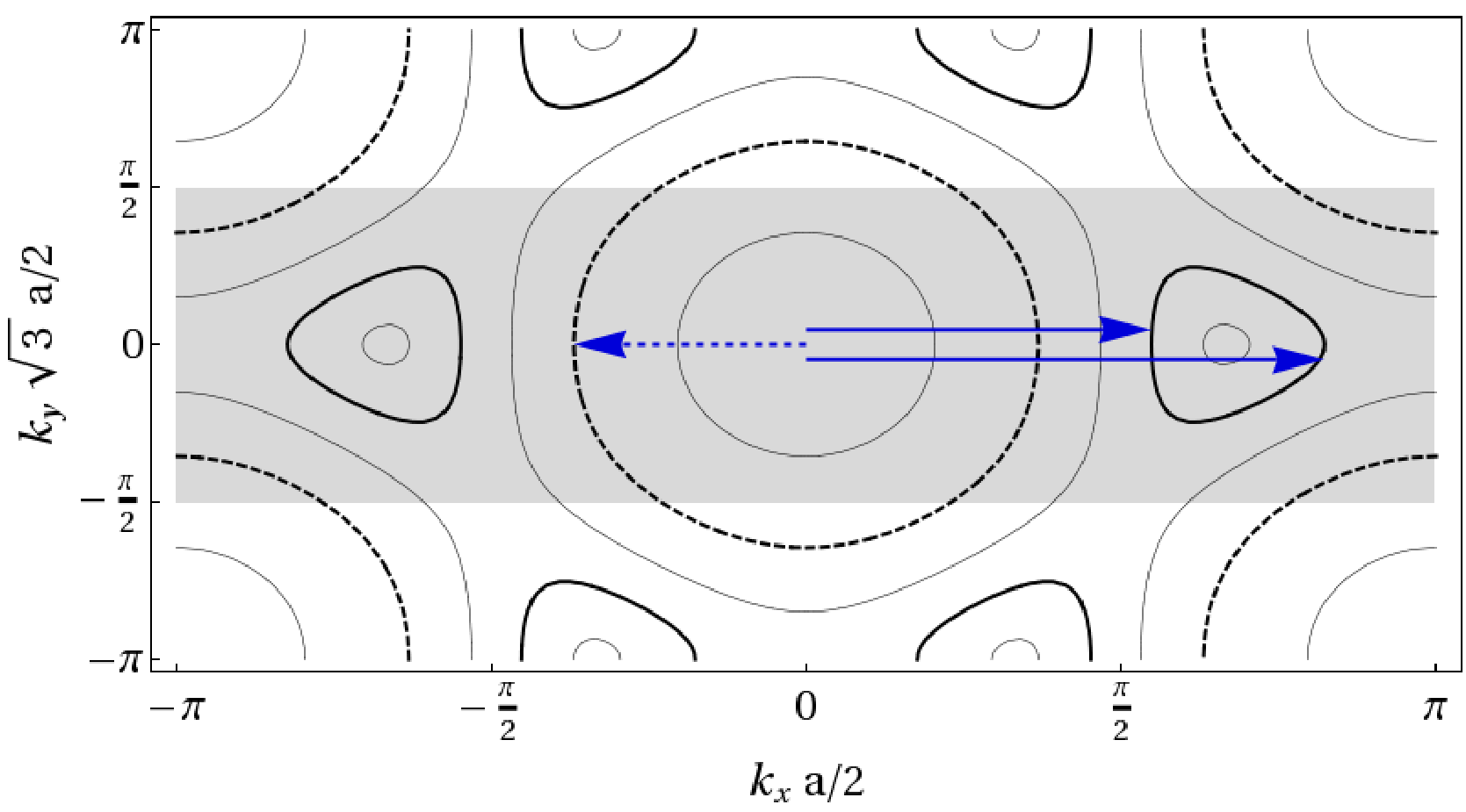}
\caption{The constant energy surfaces of the function ${\cal E}_+({\vec k})$ as before with the Brillouin Zone for the zigzag case now illustrated by the shaded area. The thick lines once more refer to constant energy plots for $E=0.7 |t|$ (solid) and $E=1.8 |t|$ (dashed). The corresponding stationary wave vectors ${\tilde q}$ for these energies are highlighted with arrows. Note that the sign of ${\tilde q}$, shown as positive (solid) or negative (dashed) only for simplicity, must be chosen according to the conditions outlined in the text. For clarity the arrows are shifted slightly in the vertical direction, but it should be noted that all the stationary points occur at $k_y=0$. }
\label{fig_zz_energies}
\end{figure}

We now turn our attention to the case of separation vectors ${\vec R}_j - {\vec R}_{j^\prime}$ along the (zigzag) $x$-direction. By following the same approach described for the armchair direction the relevant Green function can be similarly calculated. The major difference between the calculations is that the ordering of the integrals is swapped. For the zigzag direction, we first perform a contour integral over $k_x$ before making use of the SPA to solve the remaining integral over $k_y$.  We make a different choice of Brillouin Zone, as shown in Fig. \ref{fig_zz_energies}, which will simplify the selection of stationary points later. By performing the first integral similarly to before, we arrive at an expression analogous to Eq. (\ref{remaining}) for the off-diagonal Green function in the zigzag direction
\begin{equation}
{\cal G}_{\Delta}(E) = {i \, a \sqrt{3} \over 8 \pi t^2} \int_{-2\pi \over a \sqrt{3}}^{2\pi \over a \sqrt{3}} \, dk_y \, {E \, e^{i q \Delta} \over \sin({q \, a }) + \sin({q \, a \over 2 }) \cos({k_y a \sqrt{3} \over 2})}\,\,,
\label{remaining_zz}
\end{equation} 
where $q$ now represents the poles coming out of the $k_x$ integral, which are given by
\begin{equation}
\cos({q a \over 2}) = - \frac{1}{2} \left\{ \cos({k_y a \sqrt{3} \over 2}) \pm \sqrt{\frac{E^2}{t^2} - \sin^2({k_y a \sqrt{3} \over 2})} \right\} \,\,.
\label{zigzagpoles}
\end{equation}
It should be noted that in the zigzag case there are two contributions to the integral arising from the two possible sign choices of the poles in the definition above. The correct overall sign for $q$ in each case is selected as before by ensuring that $q$ lies within the integration contour. The contributions from each pole must then be summed to give the final result. As before, we assert that the only non-vanishing contributions to the integral in Eq. (\ref{remaining_zz}) occur when the phase of the exponential term is stationary. Imposing $dq/dk_y=0$ we find the stationary solution ${\tilde k}_y =  0 $.

Unlike the stationary points calculated for the armchair direction, the zigzag direction stationary points are independent of energy. This fact is evident when the constant energy plots in Fig. \ref{fig_zz_energies} are examined from the perspective of stationary values of $k_y$. The separation of the energy band into two separate regions is not necessary in this case as the stationary points for both regions occur for the same value of $k_y$. The wavevector $q$ is now Taylor expanded as before and we find expressions for $C_1$ and $C_2$ analogous to Eqs. (\ref{c1}) and (\ref{c2})

\begin{equation}
\begin{array}{l l}
C_1^{+} & = \pm {2 \over a } \, \cos^{-1}\left({t \pm E \over 2 t}\right) \\ \\
C_1^{-} & = \pm {2 \over a } \, \cos^{-1}\left({- t \pm E \over 2 t}\right) \\ \end{array} 
\label{c1_zz}  
\end{equation}
and
\begin{equation}
\begin{array}{l l}
C_2^{+} & =  \pm {t \over a E } \, \frac{t - E}{\sqrt{(3t-E)(t+E)}} \\ \\
C_2^{-} & =  \pm {t \over a E } \, \frac{t + E}{\sqrt{(3t+E)(t-E)}} \\ \end{array} \,.
\label{c2_zz}  
\end{equation}
The superscript sign in the expressions for $C_1$ and $C_2$ refer to the choice of sign in Eq. \ref{zigzagpoles}. Using these expressions the integral once more reduces to a Gaussian integral whose solution gives 

\begin{equation}
\begin{split}
{\cal G}_{\Delta}(E) =  \sum_{\alpha=\pm} \frac{i a \sqrt{3} E}{8 \pi t^2}  \sqrt{\frac{i \pi}{C_2^\alpha \Delta}} \quad \times   \quad\quad\quad\quad\quad \quad \quad \\  \quad \quad  \quad \quad \times \quad\frac{e ^ {i C_1^\alpha \Delta}} { \sin({\tilde{q}^\alpha \, a }) + \sin({\tilde{q}^\alpha \, a \over 2 }) \cos({\tilde{k_y}^\alpha a \sqrt{3} \over 2})}  \,,
\end{split}
\label{sol_zz}
\end{equation}
where the sum over $\alpha$ includes the contributions arising from the choice of sign for the poles.
Eqs. (\ref{c1_zz} - \ref{sol_zz}) can be combined as before to provide a single fully analytical expression for the off-diagonal Green function matrix element between two graphene sites separated by a distance $\Delta$ in the zigzag direction. This is found to be 

\begin{widetext}
\begin{equation}
 {\cal G}_{\Delta}(E)   = \sqrt{\frac{1}{2 i \pi \Delta^{\prime}}} \left( \sqrt{\frac{E}{|t| (t-E)}} \frac{(\frac{-t + E - i\sqrt{(3t-E)(E+t)}}{2t})^{\Delta^{\prime}}} { ((3t-E)(E+t))^{1/4} } \, + \, \sqrt{\frac{E}{|t| (t+E)}} \frac{(\frac{-t - E + i\sqrt{(3t+E)(E-t)}}{2t})^{\Delta^{\prime}}} { ((3t+E)(E-t))^{1/4} }\right) .
\label{gf_zz_et}  
\end{equation}
\end{widetext} 
where $\Delta^{\prime} = \frac{\Delta}{2 a}$. Here we have a single expression that describes the Green function across the entire band. In Fig. \ref{fig_zz_comp} we compare the expression for the Green function calculated here with the result of a  fully numerical calculation, for the case of $\Delta = 20a$. As with the armchair direction, an excellent match is found across the entire band. The plot in Fig. \ref{figerror} (triangular symbols) shows the discrepancy between the numerical and analytical results as a function of distance. Once more we find that beyond a couple of lattice spacings there is only a very narrow energy range in which the disagreement exceeds $1\%$.

\begin{figure}
\includegraphics[width=0.45\textwidth]{fig_zz_comp.eps}
\caption{${\cal G}_{\Delta}(E)$ as a function of energy (in units of $|t|$) for the case of $\Delta = 20a$. Top (Bottom) panel shows the real (imaginary) part of the Green function. Lines correspond to the results evaluated by Eq.(\ref{sol_zz}) whereas points are the result of brute-force numerical calculations of Eq.(\ref{above}). }
\label{fig_zz_comp}
\end{figure}

\subsection{Other directions and cases}
\label{oth_dir}  

Having presented the derivation of the Green function for the separation vector along the armchair and zigzag directions, it is straightforward to generalize it to other cases. For arbitrary directions, although the procedure is similar, we shall see that the identification of the poles or stationary solutions may result from high order polynomial equations that are not always analytically solvable. 

In the armchair (zigzag) case, the expression for the stationary point $\tilde{k}_x$ ($\tilde{k}_y$) is given by an easily solvable expression of the form $dq/dk=0$. This expression arises from the decision to take the contour integral along the $k$-space direction parallel to the separation vector ${\vec R}_j - {\vec R}_{j^\prime}$, which results in a phase term equal to the pole of the contour integral. Since the expressions for the poles in the armchair and zigzag directions, Eqs. (\ref{poles}) and (\ref{zigzagpoles}) respectively, are easily found from Eq. (\ref{graphene}), the calculation of all the necessary quantities is quite straightforward. To extend this approach to an arbitrary separation vector, we must first rewrite Eq. (\ref{graphene}) in terms of $k$-space vectors $k_{\parallel}$ and $k_{\perp}$ which are parallel and perpendicular respectively to the required separation vector. Following this, we must perform the contour integral over $k_{\parallel}$ to get an expression for the Green function analogous to Eq. (\ref{remaining}). However, the expression for the poles of this contour integral will depend strongly on the separation vector chosen and will usually result from a high order polynomial equation that may need to be solved numerically. It is important to note that this equation will only depend on the direction, and not the length, of the separation vector, so that once the Green function for a particular direction has been constructed it is valid for any required distance across the graphene lattice in that direction.

A similar methodology holds for the case of Green functions between sites on the different sublattices of graphene. In this case Eq. (\ref{above}) must be altered slightly to read
\begin{equation}
\langle j, \zeta \vert {\hat {\cal G}}(E) \vert j^\prime, \zeta^\prime \rangle=\frac{a^2 \sqrt{3}}{8 \pi^2 }\int dk_x \, \int dk_{y} \, \frac{t f({\vec k}) \,\,e^{i {\vec k}.({\vec R}_j - {\vec R}_{j^\prime})} }{E^{2}-{\cal E}_{+}^{2}({\vec k})}\,,
\label{Sublattices}
\end{equation}
where now we have $\zeta \ne \zeta^\prime$, with $f({\vec k}) = e^{\frac{ik_y a}{\sqrt{3}}} + 2 \,\mathrm \cos(\frac{k_x a}{2})e^{\frac{- ik_y a}{2 \sqrt{3}}}$. The integral can now be split into two parts with different phase terms coming from the two components of $f({\vec k})$. These can then be solved individually using the approach described above to give the required Green function.

\section{Application}
\label{gen_app}
The usefulness of having an analytical expression for the real space Green function, valid throughout the entire electronic energy band, becomes obvious when physical properties of graphene involving energy scales outside the linear dispersion region are investigated. This is particularly advantageous when such properties carry size or position dependence because in this case Eq.(\ref{sol_x}) for the armchair direction, or Eq.(\ref{sol_zz}) for the zigzag direction, can be more concisely expressed in the form 
\begin{equation}
{\cal G}_{\Delta}(E) = {{\cal A}(E) \, e^{i C_1(E) \Delta} \over \sqrt{\Delta}}\,\,,
\label{concise}
\end{equation}
so that the $E$- and $\Delta$-dependences are clearly distinguished. Furthermore, even in the case when the functional form of the coefficient ${\cal A}(E)$ is not particularly simple, the expressions in Eqs.(\ref{sol_x}) and (\ref{sol_zz}) can be used to expand the Green function in a polynomial series, which is undoubtedly far simpler and more treatable than the original expression in Eq.(\ref{above}). We anticipate that the ability to clearly isolate the distance dependence in the Green function will allow a more transparent treatment of some of the more eagerly investigated properties of graphene.

This approach will be shown more clearly in the next section when our formalism is used to investigate the magnetic interaction between two magnetic moments in graphene. This type of interaction is perfectly suited for investigation using our approach since it is mediated by the conduction electrons of the graphene host. However, there are many other interesting physical properties that can be explored. The interaction between precessing magnetic moments is one area of particular interest. Within the random phase approximation this dynamic coupling can be described by an integral over a complex function involving the convolution of Green functions. Initial numerical results in carbon nanotubes, which are closely related to graphene, suggest that the range of the dynamic interaction may be greater than that of the more familiar static case\cite{dynamic1, dynamic2, dynamic3}. We anticipate that an extension of the description provided below for the static case may be useful in attempting to solve the required integral analytically and understand the distance dependence of the dynamic coupling. The ability of our approach to obtain Green functions over a very large fraction of the energy band becomes increasingly important in this case due to the convolution of Green functions of different energies that appears in the expression.

Another topic that lends itself to our approach is the effect of disorder\cite{CN_disorder} and in particular, the effect of an impurity, on the properties of graphene. Friedel oscillations, occuring in the local density of states as a function of distance from an introduced impurity, have been studied within the linear dispersion regime using a Green function formalism\cite{bena-friedel}. Although the local density of states is associated with the diagonal term of the Green function, the distance dependence of the oscillations will be determined solely by the off-diagonal term calculated here. Similarly, the signatures of magnetic adatoms in graphene when probed by scanning tunneling spectroscopy have also been investigated using a theoretical approach\cite{STS}. This method again avails of Green function methods within the linear dispersion regime. We anticipate that our approach may allow for an extension of such studies to energies beyond the linear dispersion regime.

\subsection{Application to RKKY interaction}
\label{rkky_app}
To demonstrate the power and applicability of our new formalism, we turn our attention to the magnetic interaction in graphene. This interaction determines the relative orientation of magnetic moments embedded in graphene and has been the subject of many recent papers\cite{rkky_saremi, rkky_brey, rkky_voz, rkky_dugaev, rkky_bunder, rkky_blacks, rkky_sherafati}, as an understanding of this interaction is a major step in the implementation of graphene devices in the field of spintronics. When the linear dispersion approximation is used, a cut-off function is required to prevent the result diverging due to high energy contributions. There has been some debate about the effect of the cutoff function chosen on the resultant interaction calculated\cite{rkky_saremi, rkky_brey, rkky_blacks}. Other approaches to circumvent this problem involve numerical calculations\cite{rkky_blacks, rkky_sherafati} which can lack the transparency of an analytical solution. Here we shall show that the decay rate and oscillation period of the interaction as a function of distance emerge naturally from a simple calculation using our formalism and without resorting to an energy cut-off or a restriction to the linear dispersion regime. The exchange energy, $J$, within the RKKY approximation\cite{RK, K, Y} is proportional to the static susceptibility, $\chi$, which can in turn be written in terms of Green functions, allowing us to write $J_{\Delta}(E_F) \sim \mathrm{Im} \int \mathrm{d} E f(E) {\cal G}_{\Delta}^2(E) $ for two moments occupying like-sites in the graphene lattice separated by a distance $\Delta$, where $f(E)$ is the Fermi function. This quantity relates to the energy difference between the ferromagnetic and antiferromagnetic alignment of the moments, with its sign giving the energetically favourable alignment. Using the expression in Eq. (\ref{concise}) we write
\begin{equation}
 J_{\Delta} \sim \mathrm{Im} \int \mathrm{d} E \frac{{\cal B} (E) e^{2 i C_1(E) \Delta}}{\Delta (1 + e^{\beta(E - E_F)})} \,\,,
\label{J_integral}
\end{equation}
where ${\cal B} (E) = {\cal A}^{2} (E)$, $\beta = \frac{1}{k_B T}$, T being the temperature and $k_B$ the Boltzmann constant. The integral in Eq (\ref{J_integral}) can be solved by replacing it with a contour integral in the energy upper-half plane. In this case the poles are given by the zeroes of the denominator of the Fermi function, namely the Matsubara frequencies, $E_p = E_F + i (2p+1) \pi k_B T$ where $p$ is an integer labelling the poles. Writing the coefficient ${\cal B} (E)$ as a Taylor series, and the wavevector $C_1(E)$ as a first order expansion, around the Fermi energy gives
\begin{equation}
\begin{split}
 J_{\Delta} \sim \frac{k_B T}{\Delta} \sum_l \frac{1}{l !} {\cal B}^{(l)} e^{2iC_1^{(0)} \Delta} \quad \times \quad \quad \quad \quad  \\ \quad \quad \quad \quad \times \quad \sum_p e^{2iC_1^{(1)} (E_p - E_F) \Delta} (E_p - E_F)^{l}\,\,,
\label{J_sum_long}
\end{split}
\end{equation}
using ${\cal B}^{(l)}$ ($C_1^{(l)}$) to denote the $l$th derivative of ${\cal B}$ ($C_1$) evaluated at the Fermi energy.
This can be rewritten as 
\begin{equation}
 J_{\Delta} \sim \frac{1}{\Delta} \sum_l \frac{{\cal B}^{(l)}}{l !}  \frac{e^{2iC_1^{(0)} \Delta}}{(2iC_1^{(1)})^{l}} \frac{\mathrm{d}^l}{\mathrm{d} \Delta^l} \left\{\frac {k_B T}{2 \mathrm{sinh} (2 C_1^{(1)} \pi k_B T \Delta)} \right\}\,\,.
\label{J_sum_short}
\end{equation}
In the low temperature limit, $T \rightarrow 0$, this expression simplifies to one of the form
\begin{equation}
 J_{\Delta} (E_F) \sim \sum_l {\cal B}^{(l)} (E_F) \frac{e^{2i C_1 (E_F) \Delta}} {\Delta^{l+2}}\,\,.
\label{J_simple}
\end{equation}
In this form the oscillation period and decay rate of the interaction at different Fermi energies can be easily extracted. The decay rate in the asymptotic limit\cite{david2} is determined by the leading term in Eq (\ref{J_simple}), namely $l=0$, suggesting that, in general, $J \sim \Delta^{-2}$. However, at $E_F =0$, it is straightforward to determine from Eq. (\ref{gf_et}) for the armchair direction, and from Eq. (\ref{gf_zz_et}) for the zigzag direction, that the coefficient ${\cal B}^{(0)}$ vanishes and the decay rate is in fact determined by the first surviving term, $l=1$, resulting in $J \sim \Delta^{-3}$ for undoped graphene, as reported elsewhere\cite{rkky_saremi, rkky_brey, rkky_blacks, rkky_sherafati}. Also, at $E_F =0$, the oscillation period is perfectly commensurate with the graphene lattice spacing and thus oscillations are masked. When $E_F \ne 0$, the leading term does not vanish, and the oscillation period is no longer commensurate with the lattice spacing, leading to the observed oscillatory interaction\cite{rkky_brey} that decays as $J \sim \Delta^{-2}$. Note that these conclusions are reached for values of $E_F$ regardless of whether they lie within the linear dispersion regime. The correct decay rate and oscillatory behaviour for the RKKY interaction in graphene have emerged naturally and in a mathematically transparent fashion from our formalism, without resorting to the linear response approximation or the need for a cut-off function. As far as the authors are aware, this is the first time this has been performed within a fully analytical framework.

\section{Conclusions}
In summary, we have derived closed-form expressions for the single-electron Green function of graphene in real space that does not rely on the linearity of its dispersion relation near the Fermi level. The full derivation of this quantity for the two principal directions investigated on the graphene lattice has been presented along with a discussion for extending the methodology to other cases. The expressions are valid across a very large fraction of the energy band and yet remain mathematically transparent. The newly-acquired simplicity with which we can describe the electronic properties of graphene will lead to insightful new ways of studying the physical properties of this material at energy scales well beyond the linear dispersion regime. The approach described here for the magnetic interaction can be modified straightforwardly to extend the validity of expressions derived in topics including, but not limited to, modelling the dynamic magnetic coupling and the effects of adatoms in graphene systems.

\subsection*{Acknowledgments} The authors wish to acknowledge support received from Science Foundation Ireland and the Irish Research Council for Science, Engineering and Technology under the EMBARK initiative. MSF thanks A. H. Castro Neto and V. M. Pereira for fruitful discussions during his stay in Boston and the Condensed Matter Theory Visitor's Program at Boston University for support.


\begin{references}
\bibitem{hopping_note} The electronic hopping parameter of graphene is commonly quoted in the literature as approximately $t=-2.7eV$. Within this work, the energy, $E$, is expressed in units of $|t|$. However, whenever $t$ appears in an expression it is important that a negative quantity is used. It is common practice to set $t=1.0$ when dealing with expressions similar to Eqs. (\ref{gf_et}) and (\ref{gf_zz_et}). However due to the sign choices implicit in the calculation of these quantities this may lead to an erroneous result. If renormalising $t$, a negative value will return the correct Green function.
\bibitem{obs} Accounting for both stationary solutions improve the result only for energies very near the van Hove singularities at $E = \pm t$. Other than that, the excluding conditions at Eq.(\ref{tilde}) are sufficient.
\end{references}
\end{document}